\documentclass[pra,a4paper,aps,twocolumn,showpacs,superscriptaddress,groupedaddress]{revtex4}
\usepackage[dvips]{graphicx}
\usepackage{ae}
\usepackage[T1]{fontenc}
\usepackage[ansinew]{inputenc}
\usepackage{amsmath}
\usepackage{amssymb}
\usepackage{graphicx}
\usepackage{color}
\usepackage{lscape}
\usepackage[colorlinks=true, citecolor=blue, urlcolor=blue]{hyperref}  

\begin{document}
	\title{Revealing Genuine Steering under Sequential Measurement Scenario}
	
	\author{Amit Mukherjee}
	\email{amitisiphys@gmail.com }
	\affiliation{Physics and Applied Mathematics Unit, Indian Statistical Institute, 203, B. T. Road, Kolkata 700108 , India.}
	
	\author{Arup Roy}
	\affiliation{Physics and Applied Mathematics Unit, Indian Statistical Institute, 203, B. T. Road, Kolkata 700108 , India.}
	
	\author{Some Sankar Bhattacharya}
	\affiliation{Physics and Applied Mathematics Unit, Indian Statistical Institute, 203, B. T. Road, Kolkata 700108 , India.}
	
\author{Biswajit Paul}
\affiliation{Department of Mathematics, South Malda College, Malda, West Bengal, India}

\author{Kaushiki Mukherjee}
\affiliation{Department of Mathematics, Government Girls’ General Degree College, Ekbalpore, Kolkata, India.}

\author{Debasis  Sarkar}
\affiliation{Department of Applied Mathematics, University of Calcutta, 92, A.P.C. Road, Kolkata-700009, India.}

\begin{abstract}
	Genuine steering is still not well understood enough in contrast to genuine entanglement and nonlocality. Here we provide a protocol which can reveal genuine steering under some restricted operations compared to the existing witnesses of genuine multipartite steering. Our method has an impression of some sort of `hidden' protocol in the same spirit of hidden nonlocality, which is well understood in bipartite scenario. We also introduce a genuine steering measure which indicates the enhancement of genuine steering in the final state of our protocol compared to the initial states.
\end{abstract}

\maketitle

\section{I. INTRODUCTION}
Einstein-Podolsky-Rosen steering, the phenomenon that was first discussed by Schrodinger and afterwards considered as a notion of quantum nonlocality, has gained significant attention in recent days\cite{schr,wisemanpra07,wisemanprl07}. This quantum phenomenon, which has no classical analogue, is observed if one of two distant observers, sharing an entangled state, can remotely steer the particle of the other distant observer by performing measurements on his/her particle only.  The experimental criteria for analyzing the presence of  bipartite steering, first investigated in \cite{M. D. Reid09}, was formalized in ref where the authors generalized this concept for arbitrary systems\cite{wiseman11}. Till date there has been a lot of analysis regarding various features of steering nonlocality such as methods of detection\cite{Howell2013} and quantification of steering\cite{Cavalcanti2013,Aolita2014}, steering of continuous variable systems\cite{adesso15}, loop-hole free demonstration of steering\cite{zeilinger}, applications as a resource of nonlocal correlations in the field of quantum information protocols, exploiting the relation of steering with incompatibility of quantum measurements\cite{bru,ghu} and its ability to detect bound entanglement\cite{ghune14}, etc. Apart from its foundational richness, EPR steering do have multi-faceted applications in practical tasks such as semi-device independent scenario \cite{wiseman2013} where only one party can trust his or her apparatus but the other party's apparatus is not trusted. In that situation the presence of steerable state provides a better chance to allow secure key distribution\cite{branciard2012}. Even for some other tasks such as randomness certification\cite{acin15}, entanglement  assisted  sub-channel  discrimination\cite{watrous}, and secure teleportation through continuous-variables steerable states\cite{reid15} are found to be useful.\\
Being a notion of nonlocality there exists a hierarchy according to which steering is defined as a form of quantum inseparability, intermediate in between entanglement and Bell nonlocality. Considering pure quantum states these three notions are equivalent whereas in general they are inequivalent in case of mixed states \cite{brunner15}. However in the context of comparison of steering nonlocality with that of Bell-nonlocality, it is interesting to mention that analogous to hidden nonlocality\cite{popescu95,gisin96}, existence of hidden steering has been proved in \cite{brunner15} for bi-partite scenario. Just as in the case of exploiting nonlocality beyond Bell scenarios via the notion of hidden nonlocality\cite{popescu95,gisin96}, hidden steering refers to revelation of steering nonlocality under suitable sequential measurements. In this context an obvious interest grows regarding analysis of the same for multipartite scenario.\\
Due to increase in complexity as one shifts from bipartite to  multipartite system, till date there has been limited attempts to understand the feature of multipartite steering phenomenon. Analogous to both entanglement and Bell-nonlocality the concept of genuine steering has been established in recent days. In this context it may be mentioned that unlike Bell-nonlocality and entanglement, due to asymmetric nature of steering nonlocality the notion of genuine steering nonlocality lacks uniqueness. However  genuine steering was first introduced in \cite{reid} where the authors provided the criteria for detecting genuineness in steering scenario  for both continuous as well as discrete variable systems.  Later two other notions of genuine steering were introduced in \cite{jeeba} mainly for tripartite framework where two parties measurements are fully specified i.e one party can control remaining. In this context, the author has also designed genuine steering inequalities to detect genuine tripartite steering. Now speaking of genuine steering nonlocality, it may be interesting to explore the possibility of exploiting the same via some suitable sequential measurement protocol. \\
To be precise, our present topic of discussion will continue in the direction of analyzing hidden genuine tripartite steering nonlocality in the framework introduced in \cite{jeeba}. For present topic of discussion we will follow terms and terminologies used in \cite{jeeba}. We will design a protocol involving a sequence of measurements such that initially starting from tripartite states which may not be genuinely steerable, the protocol may generate a genuinely steerable state. Interestingly the initial states which will be used in the protocol do have a bilocal model \cite{acin2014}.\\

The paper has organized as follows. In section[II] we have introduced the notion of steering both in bi partite as well as genuine multipartite scenario. Then in section[III] we have presented suitable sequential operations to achieve the final state. Section[IV] contains our main results then discussion.

 \section{BACKGROUND}
 In this section we are basically going to include a brief detailing of the mathematical tools that will be used in our work.

\subsection{Genuine tripartite steering}
Firstly we discuss the criteria of detecting genuine steering\cite{jeeba}. Correlations $P(a,b,c|x,y,z)$  shared between three parties, say Alice, Bob and Charlie are said to be genuinely steerable\cite{jeeba} from one party, say Charlie to remaining two parties Alice and Bob, if those are inexplicable in the following form:
\begin{eqnarray}\label{jeva1}
   &&P(a,b,c|x,y,z)= \sum_{\lambda}q_\lambda [P(a,b|x,y,\rho_{AB}(\lambda))]P(c|z,\lambda)\nonumber\\
   &+&\sum_{\lambda}p_{\lambda}P(a|x,\rho^\lambda_a)P(b|y,\rho^\lambda_b)P(c|z,\lambda).
\end{eqnarray}

where $P(a,b|x,y,\rho_{AB}(\lambda))$ denotes the nonlocal probability distribution arising from two-qubit state $\rho^\lambda_{AB}$, and $P(a|x, \rho^\lambda _{A})$ and
$P(b|y, \rho^\lambda_{B})$ are the distributions arising from qubit states $\rho^\lambda _{A}$
and $\rho^\lambda_B$.

Here Charlie performs uncharacterized measurement whereas both Alice and Bob have access to qubit measurements. The tripartite correlation will be called genuinely unsteerable if it is explained by \ref{jeva1} where $\rho_{AB}(\lambda)$ is called hidden state for Alice and Bob side.
In \cite{jeeba}, the author designed a detection criteria of tripartite genuine steering(Svetlichny steering), based on Svetlichny inequality\cite{SI}. The detection criterion is given in the form of a Bell-type inequality:
\begin{equation}\label{pst5}
\langle CHSH_{AB}z_1+CHSH^{'}_{AB}z_0\rangle_{2\times2\times ?}^{NLHS}\leq 2\sqrt{2}.
\end{equation}
where $CHSH_{AB}$ and $CHSH^{'}_{AB}$ stand for two inequivalent facets defining Bell-CHSH polytope for Alice and Bob and $\{z_0,z_1\}$ are measurements on Charlie's part. Here $NLHS$ stands for nonlocal hidden state whereas $2\times2\times ?$ implies that only two parties(Alice and Bob) have access to qubit measurements but Charlie does not trust his measurement devices and hence are uncharacterized. Alice and bob should have orthonormal measurement settings. If correlations arising due to measurements on any given quantum state($\rho$) violate this inequality(Eq.(\ref{pst5})), then that guarantees genuinely steerable of $\rho$ from Charlie to Alice and Bob. Analogously genuine steerability of $\rho$ from Bob to Charlie and Alice and that from Alice to Charlie and Bob can be guaranteed respectively by violation of the following criteria:
\begin{equation}\label{pst5a}
\langle CHSH_{BC}x_1+CHSH^{'}_{BC}x_0\rangle_{2\times2\times ?}^{NLHS}\leq 2\sqrt{2}.
\end{equation}
\begin{equation}\label{pst5b}
\langle CHSH_{AC}y_1+CHSH^{'}_{AC}y_0\rangle_{2\times2\times ?}^{NLHS}\leq 2\sqrt{2}.
\end{equation}
Terms $CHSH_{BC}$, $CHSH^{'}_{BC}$, $CHSH_{AC}$, $CHSH^{'}_{AC}$ have analogous definitions. Hence a state is genuinely steerable from one party to the remaining two parties if it can violate atleast one of these three criteria(Eqs.\ref{pst5},\ref{pst5a},\ref{pst5b}).\\
We now discuss about some relevant tools for measuring genuine multipartite entanglement and genuine steering.
\subsection{Genuine multipartite concurrence}
We briefly now describe $C_{GM}$, a measure of genuine multipartite entanglement. For pure $n$-partite states($|\psi\rangle$), this measure is defined as \cite{Zma} :
$C_{GM}(|\psi\rangle):= \textmd{min}_j\sqrt{2 (1-\Pi_j(|\psi\rangle))}$ where $\Pi_j(|\psi\rangle)$ is the purity of $j^{th}$ bipartition of $|\psi\rangle$. The expression of $C_{GM}$ for $X$ states is given in \cite{Has}. For tripartite $X$ states,
\begin{equation}\label{4v}
C_{GM}=2\,\textmd{max}_i\{0,|\gamma_i|-w_i\}
\end{equation}
with $w_i=\sum_{j\neq i}\sqrt{a_jb_j}$ where $a_j$, $b_j$ and $\gamma_j(j=1,2,3,4)$ are the elements of the density matrix of tripartite X state:

\[\begin{bmatrix}
a_1 & 0    & 0    & 0   &  0   &  0    & 0    & \gamma_1 \\
0   & a_2  & 0    & 0   &  0   &  0    & \gamma_2  &  0  \\
0   & 0    & a_3  & 0   &  0   &  \gamma_3  & 0    &  0  \\
0   & 0    & 0    & a_4 &  \gamma_4 &  0    & 0    &  0  \\
0   & 0    & 0   &{\gamma_4}^\ast &b_4 & 0  & 0   &    0   \\
0   & 0    &{\gamma_3}^\ast   &0 &  & b_3  & 0   &    0  \\
0   &{\gamma_2}^\ast    & 0  & 0 & 0 & 0 & b_2 & 0 \\
{\gamma_1}^\ast & 0 &0 & 0 & 0 & 0 & 0 & b_1 \\
\end{bmatrix}\]
\subsection{Genuine steering measure}
First we define genuine steering measure which is analogous to the bi-partite steering measure first described in \cite{angelo16}. This measure is given by the following quantity:
\begin{equation}\label{sm}
 S_{gen}(\rho)= max\{0,\frac{S_n(\rho)-1}{S_n^{max}-1}\}
\end{equation}
where$S^{max}_{n}=\max_\rho S_{n}(\rho)$ and $S_n(\rho)=\max_\eta S_n(\rho,\eta)$ with the maximization taken over all measurement settings $\eta$ and $0\le S_{gen}(\rho)\le 1$.\\
After giving a brief detailing of our mathematical tools, we now proceed with our results. To start with, we design the sequential measurement protocol based on which we observe the enhanced revelation of genuine steering.  
\section{Revealing Multipartite Genuine Steering}\label{proto}

\begin{figure}[b!]
	\includegraphics[scale=0.8]{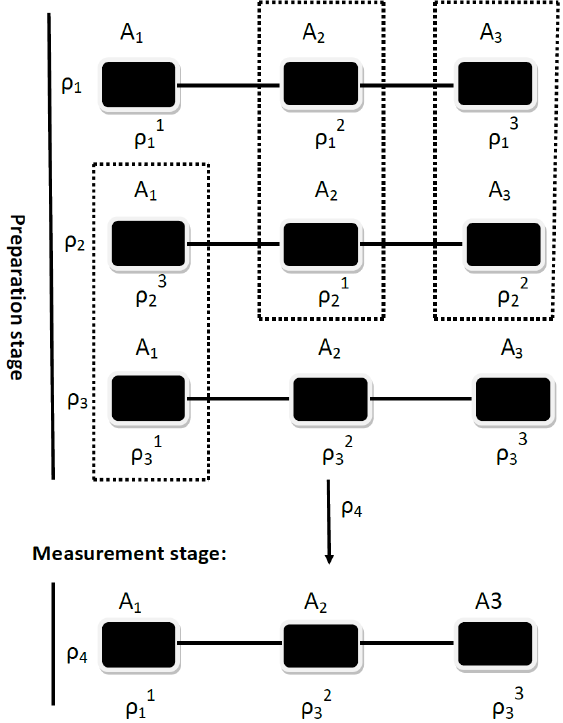}
	\caption{Schematic diagram for preparation and mesurement stage.}\label{sets}
\end{figure}

The protocol that we propose here is a SLOCC(Stochastic Local Operation and Classical Communication) protocol which consists of two stages: \textit{Preparation Stage} and \textit{Measurement Stage}. We name this protocol as \textit{Sequential Measurement Protocol}. A detailed sketch of the protocol is given below:\\
\textit{Sequential Measurement Protocol}: 
 Three spatially separated parties(say, $A_i;i=1,2,3$) are involved in this protocol. $n$ number of tripartite quantum states can be distributed among them. None of these states violate genuine steering inequality\cite{jeeba}. As each party holds one particle from each of the n tripartite states hence each of the parties holds n number of particles. 
 \subsection{Preparation Stage} 
 \begin{itemize}
 	\item In the preparation stage, every party can perform some joint measurement on their respective $n-1$ particles and then broadcast the results to others.
 	
 	\item At the end of measurements by all the three parties, a tripartite quantum state shared among $A_1$, $A_2$ and $A_3$ is generated. Clearly this final state is always prepared depending upon the measurement results obtained by the parties in the previous step.
 \end{itemize}
    \subsection{Measurement Stage} 
    \begin{itemize}
    	\item In the measurement stage, all the three parties can perform any projective measurement in arbitrary directions. But in this stage they are not allowed to communicate among themselves.   
    	
    	\item After measurements they can generate a tripartite correlation so that they can verify that this correlation can violate the genuine steering inequality.
    \end{itemize}
We refer to this protocol of sequential measurements by the three parties sharing n states as a sequential measurement protocol (SMP).

Having sketched the protocol we now give examples of some families of tripartite states which when used in this network, reveal genuine steering for some members of these families. Such an observation is supported with an increase in the amount of genuine steering, guaranteed by the measure of steering $S_{gen}(\rho)$(Eq.(\ref{sm})).\\

Let the three initial states be given by:
\begin{equation}\label{1}
\rho_1 = p_1 |\psi_f\rangle\langle \psi_f|+(1-
p_1)|001\rangle\langle001|
\end{equation}
with $|\psi_f\rangle=\cos\theta_1|000\rangle+\sin\theta_1|111\rangle$, $0\leq\theta_1\leq \frac{\pi}{4}$ and $0\leq p_1\leq 1$;
\begin{equation}\label{2}
\rho_2=  p_2 |\psi_m\rangle\langle \psi_m|+(1-p_2)|010\rangle\langle010|
\end{equation}
with $|\psi_m\rangle=\frac{|000\rangle+|111\rangle}{\sqrt{2}}$ and $0\leq p_2\leq 1$;
\begin{equation}\label{3}
\rho_3=  p_3 |\psi_l\rangle\langle \psi_l|+(1-p_3)|100\rangle\langle100|
\end{equation}
with $|\psi_l\rangle=\sin\theta_3|000\rangle+\cos\theta_3|111\rangle$,$0\leq\theta_1\leq \frac{\pi}{4}$ and $0\leq p_1\leq 1$.
In this context it may be noted that the three initial states have Svetlichny bi-local model under projective measurement for the following restricted range of state parameters:
\begin{itemize}
\item For first state($\rho_1$) : $p_1 \leq \frac{1}{(1 + \sin[2\theta_1])}$;
\item Second state($\rho_2$) : $p_2 \leq \frac{1}{2}$ ;
\item Third state($\rho_3$) : $p_3 \leq \frac{1}{(1 + \sin[2\theta_3])}$.
\end{itemize}
Each of the three parties $A_1$, $A_2$ and $A_3$ performs Bell basis measurements on their respective particles. Depending on a particular output of all the measurements(here $|\psi^{\pm}\rangle=\frac{|01\rangle\pm|10\rangle}{\sqrt{2}}$), a resultant state $\rho_4^{\pm}$ is obtained which after correcting the phase term is given by:
  \begin{equation}\label{4}
    \rho_4 = \frac{p_3 |\phi\rangle\langle\phi| + (1 - p_3)\sin^2 \theta_1|100\rangle\langle100|}{\sin^2 \theta_1+ p_3 \cos2\theta_1 \sin^2 \theta_3 }
  \end{equation}
  where $|\phi\rangle = \cos\theta_1 \sin\theta_3|000\rangle + \sin\theta_1 \cos\theta_3 |111\rangle$.

 Clearly $\rho_4$ is independent of $p_1$ and $p_2$. Interestingly, $\rho_4$ can also be generated for some other combination of sequential operations on some different arrangement of particles between the parties $A_i(1,2,3)$ and for different output of Bell measurement.
For the initial states $\rho_i$ $(i = 1, 2, 3)$, the amount of genuine entanglement are given by
$$ C^{\rho_1}_{GM} = p_1 \sin 2\theta_1 ,$$
$$ C^{\rho_2}_{GM} = p_2 $$ and
\begin{equation}\label{6iii}
C^{\rho_3}_{GM} = p_3 \sin 2\theta_3
\end{equation}
whereas that of $\rho_4$ is given by
\begin{equation}\label{6iv}
C^{\rho_4}_{GM} = \frac{p_3 \sin2\theta_1\sin2\theta_3 }{2(\sin^2 \theta_1+ p_3 \cos2\theta_1 \sin^2 \theta_3)}.
\end{equation}
Eq.(\ref{6iii}) indicates that the initial states $\rho_i(i=1,2,3)$ are  genuinely entangled for any nonzero value of the state parameters .

The maximum value of the genuine steering operators($S_i$)(Eq.\ref{pst5}) under projective measurements, for state $\rho_i(i=1,2,3)$ is given by:
\begin{widetext}
$$S_1 = \max[2\, p_1\sin2\theta_1, \frac{1}{\sqrt{2}}\sqrt{((1-p_1-p_1 Cos[2 \theta_1])^2+(p_1 Sin[2\theta_1])^2}],$$
\end{widetext}
 
  $$S_2 = \max[2\, p_2,  \frac{1}{\sqrt{2}}\sqrt{((1-p_2)^2+(p_2)^2)}]\,\,\,\,$$ and
 \begin{equation}\label{6i}
S_3 = \max[2\, p_3\sin2\theta_3, \frac{1}{\sqrt{2}}\sqrt{((1-p_3+p_3 Cos[2 \theta_3])^2+(p_3 Sin[2\theta_3])^2} )|]
\end{equation}
respectively whereas that for the final state $\rho_4$, it is given by
\begin{widetext}
$$S_4 = \max[\frac{\,p_3 \sin2\theta_1\sin2\theta_3}{\sin^2 \theta_1+ p_3 \cos2\theta_1 \sin^2 \theta_3},$$
\begin{equation}\label{6ii}
\frac{\sqrt{2} \sqrt{(1-p_3+p_3 Cos[2 \theta_3]-Cos[2 \theta_1])^2+(p_3 Sin[2 \theta_1] Sin[2\theta_3])^2}}{2-2(1-p_3)Cos[2 \theta_1]-p_3Cos[2(\theta_1-\theta_3)]-p_3 Cos[2(\theta_1+\theta_3)]}.
\end{equation}
\end{widetext}

It is clear from the maximum  value of genuine steering operator(Eqs.(\ref{6i}), (\ref{6ii})) and the measure of entanglement (Eqs.(\ref{6iii}), (\ref{6iv})) of both initial states and final state, that each of them does not violate genuine steering inequalities(Eqs.\ref{pst5},\ref{pst5a},\ref{pst5b}) for $C^{\rho_i}_{GM}\leq \frac{1}{2}(i=1,2,3,4).$
\par
Thus to observe genuine steering revelation there should exist some fixed values of the parameters of the three initial Sveltlichny bi-local states with $C^{\rho_i}_{GM}\leq\frac{1}{2}$ such that the final state can have $C^{\rho_4}_{GM}>\frac{1}{2}$. Interestingly we get such states from the families of the initial states $\rho_1$(Eq.(\ref{1})), $\rho_2$(Eq.(\ref{2})) and $\rho_3$(Eq.(\ref{3})).
\par
For example, let $\theta_1=0.1$, $p_1\leq 0.509 $ , $p_2\leq\frac{1}{2}$, $\theta_3=0.1$ and $p_3\in[0,0.83426]$. Then each of the initial states have Svetlichny bi-local model (moreover one can show that these models are $NS_2$ local\cite{acin2014}) and $C^{\rho_i}_{GM}\leq\frac{1}{2}$. Thus they do not violate genuine steering inequalities(Eqs.\ref{pst5},\ref{pst5a},\ref{pst5b}).
\par
But when used in our protocol(Sec.\ref{proto}), they can generate a state $\rho_4$ (with $C^{\rho_4}_{GM}>\frac{1}{2}$) which exhibits genuine steering by violating genuine steering inequalities for $p_3 \geq 0.33557$. This guarantees revelation of genuine steering for $p_3 \in [0.33557,0.83426]$. So initially each of these three states are unable to exhibit genuine steering but after the sequential measurements are taken into account they can violate that genuine steering inequality. Now a pertinent question would be whether one can quantify this revelation of genuine steering as observed in our protocol. We deal with this question in the next sub-section. 
 \subsection{Enhancement of the Genuine Steering measure}
 In this part we show that the prescribed protocol indeed enhances a measure of genuine steering in the resulting state. The amount of genuine steering for the three initial states are:
 $$S_{gen}(\rho_1)= max\{0,2\, p_1\sin2\theta_1-1\},$$
  $$S_{gen}(\rho_2)= max\{0,2\, p_2-1\},$$
  \begin{equation}
S_{gen}(\rho_3)= max\{0,2\, p_3\sin2\theta_3-1\}
  \end{equation}
whereas for the final states the genuine steerable quantity takes the form:
  \begin{equation}
S_{gen}(\rho_4)= max\{0,\frac{p_3 \sin2\theta_1\sin2\theta_3}{\sin^2 \theta_1+ p_3 \cos2\theta_1 \sin^2 \theta_3}-1\}
  \end{equation}
If we take $p_1=p_3$ and $\theta_1=\theta_3$ then for any values of $p_1$ and $\theta_1$ the final state is more genuinely steerable than the initial ones.
\section{CONCLUSION}
Genuine steering nonlocality, being a weaker notion of genuine nonlocality is considered to be a resource in various practical tasks. So apart from its theoretical importance, revelation of such a resource under any protocol that allows only classical communication and shared randomness is of immense practical importance. Motivated by that we have attempted to design a SLOCC protocol which demonstrates revelation of `hidden' genuine steering. Our discussion in a restricted sense guarantees the fact that under suitable measurements by the parties involved in the network, our protocol is sufficient to show genuine steering even from some quantum states which have bi-local models. However under our protocol each of the parties having two particles perform  Bell basis measurements and the remaining parties perform projective measurements. In brief, this protocol enables one to go beyond the scope of existing witnesses of genuine steering and thus demonstrate genuine steering for a larger class of multipartite states. In this context, it will be interesting to consider more generalized measurement settings by the parties which may be yielding better results. 
\section{Acknowledgement}

We would like to thank Prof. Guruprasad Kar for useful discussions. AM acknowledge support from the CSIR project 09/093(0148)/2012-EMR-I.

\appendix
\end{document}